\documentclass[preprint,showkeys,aps,prb]{revtex4}
\usepackage[dvips]{graphicx}
\begin{document}

\title{Aspects of Dynamical Simulations,\\
 Emphasizing Nos\'e and Nos\'e-Hoover Dynamics\\
and the Compressible Baker Map}

\author{William Graham Hoover and Carol Griswold Hoover,\\ Ruby Valley Research Institute,
601 Highway Contract 60,\\ Ruby Valley, Nevada 89833}

\date{\today}

\vspace{0.1cm}

\begin{abstract}
Aspects of the Nos\'e and Nos\'e-Hoover dynamics developed in 1983-1984 along with Dettmann's
closely related dynamics of 1996, are considered. We emphasize paradoxes associated with Liouville's
Theorem. Our account is
pedagogical, focused on the harmonic oscillator for simplicity, though exactly the same ideas can
be, and have been, applied to manybody systems. Nos\'e, Nos\'e-Hoover, and Dettmann flows were
all developed in order to access Gibbs' canonical ensemble directly from molecular dynamics.
Unlike Monte Carlo algorithms dynamical flow models are often not ergodic and so can fail to
reproduce Gibbs' ensembles.  Accordingly we include a discussion of ergodicity, the visiting of
all relevant microstates corresponding to the desired ensemble. We consider Lyapunov instability
too, the usual mechanism for phase-space mixing.  We show that thermostated harmonic
oscillator dynamics can be simultaneously expanding, incompressible, or contracting, depending
upon the chosen ``phase space''. The fractal nature of nonequilibrium flows is also illustrated for
two simple two-dimensional models, the hard-disk-based Galton Board and the time-reversible
Baker Map. The simultaneous treatment of flows as one-dimensional and many-dimensional suggests
some interesting topological problems for future investigations.
\end{abstract}

\maketitle

\newpage

\section{Introduction: A Summary of Dynamical Milestones}
Gibbs' formulation of statistical mechanics uses averages over all of phase space as predictors
of equilibrium behavior. Metropolis, the Rosenbluths, and the Tellers implemented Gibbs' program
with their ``Monte Carlo'' algorithm\cite{b1}.  They emphasized that their algorithm is ergodic,
capable of covering all of phase space, and applied it successfully to the equation of state of
hard disks. Gibbs showed that phase-space averages over ``all states'' could be used to calculate
properties of systems at constant energy, or, by weighting the states differently, at constant
temperature or pressure or both.  Some 50 years later, computers made it possible to compute such
averages numerically, at first for very small systems. This was accomplished by moving particles
with probabilities corresponding to the intended ensemble\cite{b1}. We provide two educational
examples of Monte Carlo simulation in Section II.

Berni Alder and Tom Wainwright developed ``Molecular Dynamics'' shortly thereafter\cite{b2}.  The
two methods agreed well for fixed-energy hard-sphere systems at equilibrium. Fixed-temperature
Monte Carlo simulation is limited to equilibrium problems.  Equilibrium Monte Carlo simulations
had no precise dynamical analog until Nos\'e's innovative work of 1984\cite{b3,b4,b5,b6}. 
``Nonequilibrium Molecular Dynamics'', which could deal with flows of momentum and energy, had
been developed in the 1970s\cite{b7}. The velocities, kinetic temperatures, and heat fluxes could
all be controlled by brute-force adjusting and rescaling. The resulting fixed moments of the
distribution turned out to promote viscous and thermal flows in line with hydrodynamic expectations
in relatively small systems, with only a few or at most a few hundred interacting particles. We
describe some equilibrium example flows in Section III. It was discovered that the brute-force
rescaling was equivalent to linear feedback control of the momenta in the differential equations
governing particle motion.  We illustrate this equivalence in Section IV and introduce the Galton
Board problem as the prototype example.

In 1984 Shuichi Nos\'e formulated a particle mechanics consistent with Gibbs' canonical ensemble.
His work was grounded in Hamiltonian mechanics but with a novel addition designed to promote energy
mixing. Nos\'e's mechanism for mixing states of different energies was an external ``time-scaling''
variable $s$.  A dozen years later Dettmann discovered that $s$ can best be interpreted as the
phase-space probability density, $s = f(q,p,\zeta)$, where the ``friction coefficient'' $\zeta$ in
Nos\'e's work is the momentum conjugate to $s$, $p_s = \zeta$.  This same notion was independently
discovered three years later by Bond, Leimkuhler, and Laird\cite{b8}.  Their work attracted Nos\'e's
interest more strongly than had Dettmann's \cite{b7,b9,b10}, which Bill had discussed with Nos\'e in
1996. Nos\'e's 1984 work and Dettmann's 1996 achievement are detailed in Section V.

Meeting Nos\'e in Paris in 1984 Bill Hoover developed a flow-based model, like Liouville's Theorem
for incompressible Hamiltonian flows, but applying to compressible flows of just the kind invented
by Nos\'e. Hoover emphasized the usefulness of the harmonic oscillator in understanding Nos\'e's
approach\cite{b5}. The simplicity of that model led to its independent rediscovery by Sprott a
decade later\cite{b11,b12}.  We adopt that same model here as our primary pedagogical tool. The
simplest possible thermostating mechanism, linear feedback, gave rise to a Gaussian distribution
for the friction coefficient $\zeta=p_s$, as is described in Section V.

Thermostating from a more general viewpoint, using higher moments, was glimpsed by Hoover in
1985\cite{b13}, and treated comprehensively in a pair of important papers\cite{b14} by Bauer, Bulgac,
and Kusnezov in the early 90s. The 2015 (Ian) Snook Prize problem led Tapias, Bravetti, and
Sanders\cite{b15} to a ``Logistic'' thermostat with an arctangent switch between the heating and
cooling functions of their external frictional control variable. Their work was soon elaborated by
Sprott\cite{b16}, who considered on-off ``bang-bang'' control of temperature and showed numerically
that the resulting control variable has a singular exponential distribution in this limit rather
than the Gaussian distribution resulting from Nos\'e's integral control. These ideas are also
illustrated in Section V.

In Section VI we illustrate and discuss a paradox involving three descriptions of the Nos\'e-Hoover
oscillator flow. These suggest (wrongly, of course) that the same flow can be simultaneously expanding,
contracting, or incompressible, depending upon the coordinates used to describe the flow.

Section VII provides our views on numerical methods, developed over 40 years, and influenced
particularly by our colleague Clint Sprott\cite{b17}, whose imaginative use of color has provided a
powerful adjunct to the understanding of nonlinear flows. Because bifurcation and chaos is necessary
to any statistical view of dynamical systems we include the characterization of flows in terms of
their Lyapunov spectra in this Section.

Sections VIII and IX deal with our general conclusions concerning the simulation of nonequilibrium
steady states, the formation of fractal repellors and attractors and the association of the
macroscopic Second Law of Thermodynamics with microscopic thermostated models.

Our final Section X is a bit speculative, as must be any view of the future, and suggests that the
fractal geometry of nonlinear chaotic flows still holds more interesting lessons for us.  We stress
the difference between the continuous, paradoxical, and contentious view of mathematics and the
discrete grid-based approach of computational statistical mechanics.

\newpage

\section{Monte Carlo as developed in the 1950s}

As computer hardware improved in the American National Laboratories, predominantly at Los Alamos and Livermore, the
applications became scientifically interesting. In the decade following the Second World War both
Enrico Fermi and Edward Teller became interested in developing computer applications to problems
with statistical mechanical roots. For the first time manybody problems and problems involving
billions of operations became tractable.  The Monte Carlo technique\cite{b1} was developed at
the Los Alamos laboratory, with equation of state results calculated in the early 1950s followed
soon after by dynamical studies of one-dimensional anharmonic chains, the ``Fermi-Pasta-Ulam
problem'', by 1953\cite{b18}. That innovative work was soon followed by Alder and Wainwright's
dynamical studies of hard disks and spheres\cite{b19} complemented by Wood and Jacobsen's Monte Carlo
simulations of these same systems\cite{b20}. Let us next look at two illustrative examples of the Monte
Carlo simulation technique with two simple systems, the harmonic oscillator and its quartic-oscillator relative.

\subsection{Monte Carlo Evaluation of Canonical Harmonic Oscillator Moments}

As a warmup demonstration problem let us apply the Monte Carlo method to a test problem with well-known
dynamical and ensemble-averaged answers, the one-dimensional harmonic oscillator.  As usual we choose
the mass, force constant, temperature, and Boltznann's constant all equal to unity. Gibbs' canonical
phase-space distribution for the oscillator coordinate-momentum $(q,p)$ pair, is the simple Gaussian :
$f(q,p) = e^{-(q^2 + p^2)/2}/(2\pi)$. The resulting mean values of the even moments of $q$ and $p$ are
products of the odd integers :
$$
\langle \ q^2,p^2,q^4,p^4,q^6,p^6,q^8,p^8 \dots \ \rangle = 1, 1,3,3,15,15,105,105, \dots \ .
$$

Metropolis' group published the method in 1953\cite{b1}--(and there is some controversy as to the relative
contributions of the five researchers)\cite{b18}.  They pointed out that a detailed balance between two
states of energy $E_i,E_j$ with relative probabilities obeying Gibbs' canonical (exponential) distribution,
$f_i/f_j = e^{-E_i+E_j}$, can be achieved by an imaginary equilibrium dynamics, a ``Monte Carlo simulation'',
in which changes of state occur at a definite rate.  Consider just one pair of energy states.  If the
transition rate from the lower state (say the $i$th state) is less than that from the higher state by a factor
$e^{E_j-E_i}$, just offsetting the relative probabilities of the states, a stationary state results. This
state has the desired canonical ratio, $f_i/f_j = e^{-E_i+E_j}$. If this can be achieved for all pairs of
states, and all such pairs are accessible, then Gibbs' canonical ensemble can be realized numerically, as a
limiting case. As the number of states is astronomical, for even a one-body system, it is necessary to
understand the convergence rate of Monte Carlo simulations.  Let us pursue the oscillator problem with its
known canonical distribution of the coordinate, $e^{-q^2/2}/\sqrt{2\pi}$.

A Monte Carlo program implementing this idea for the oscillator coordinate causes a single test oscillator to
make random jumps within a spatial interval $-J<dq<+J$.  The jump $dq$ occurs with probability 1 if the
energy drops and with a lesser probability $e^{-\delta E}$ if it rises.  The uphill jump with probability
$e^{-\delta E}$ is implemented by choosing an additional random number $0 < {\cal R} < 1$ and accepting the move
when ${\cal R}$ is sufficiently small, ${\cal R} < e^{-\delta E}$. This single-particle oscillator program needs
two random numbers when the new energy is higher---one for the jump and one for the acceptance test.  Only one
random number is needed (for the jump alone) when the energy is lower. The heart of the program can be summarized
by a single line of pseudocode :\\

{\tt
if((Enew.lt.Eold).or.(rund(intx,inty).lt.dexp(-Enew+Eold))) qold = qnew \\
}

We wrote such a billion-jump program using the following simple generator {\tt rund(intx,inty)}, where the two arguments
are the ``seeds'' of the random number ${\tt rund}$.  As the routine is called the corresponding sequence of {\tt intx}
and {\tt inty} values goes through all 4,194,304 combinations $0\le{\tt (intx,inty)}\le 2047$. We began with
{\tt (intx,inty)} both zero. These seeds change each time a new number is generated.  Here is {\tt rund} :\\
{\tt
\noindent
      i = 1029*intx + 1731                     \\
      j = i + 1029*inty + 507*intx - 1731      \\
      intx = mod(i,2048)                       \\
      j = j + (i - intx)/2048                  \\
      inty = mod(j,2048)                       \\
      rund = (intx + 2048*inty)/4194304.d00    \\
}
This choice reproduces the second and fourth moments $\langle \ q^2,q^4 \ \rangle$ within 0.01 for maximum jump lengths
$J$ of 1, 2, or 4, where $[-J<dq<+J]$. \ The mean squared jump length $\langle \ dq^2 \ \rangle$  was 0.61 for $J=2$,
0.87 for $J=4$, and 0.53 for $J=8$, suggesting that the relatively large jump-length interval with $J=4$ is best from
the standpoint of phase-space exploration.

A longer run of one billion Monte Carlo steps with $J=1/2$ revealed an interesting oscillation in the fourth
moment, $\langle \ q^4 \ \rangle$, with a periodic duration of roughly 30 million steps. The long-time mean
value of that fourth  moment appears to be converging in the neighborhood of 2.983 rather than exactly 3. The
relatively short period and the half-percent error in the moment suggest that the {\tt rund} generator is not
well suited to this type of Monte Carlo simulation.

The built-in {\tt gfortran} generator, {\tt rand(intx)} is arguably better, but still far from perfect. With
{\tt rand} a ten-billion step run reveals a period on the order of one billion steps converging in the
neighborhood of $\langle \ q^4 \ \rangle = 2.9984$. A similar run, but discarding the first $10^8$ random numbers,
leads to a similar period with apparent convergence to 2.9985. Finally, a program using {\it pairs} of random
numbers for each step (an effort to enhance and better characterize periodicity) gave apparent convergence to
3.00007, again with oscillation periods of about one billion steps. A careful look revealed that the sequence of
random numbers produced by {\tt rand} repeats precisely after 715,827,882 calls and evidently creates a resonant
periodicity, with that same frequency, in the oscillator itself.

This same strategy, accepting moves raising the potential energy with relative probability $e^{-\delta E}$
has been successfully applied to manybody problems ever since Metropolis' work. Such Monte Carlo sampling
is not extendable to ``nonequilibrium'' simulations (those with specified velocity or temperature gradients
for example) while Nos\'e's method is\cite{b21}. By simulating random jumps in phase space the Monte Carlo
approach automatically accesses configurations over a wide range of potential energies.

This harmonic oscillator example teaches an important lesson: test random number implementations with a few
simple applications having known answers prior to embarking on a ``new'' type of simulation.  In Section VIII
we will use ``{\tt Random\_Number(r)}'', a better FORTRAN random number generator, in another Monte Carlo
application.

\subsection{Monte Carlo Construction of Quartic Oscillator Ensembles}

Another application of the Monte Carlo method is the construction of small-system Gibbs' ensembles for given
values of the energy or temperature.  Although dynamical techniques able to generate such an ensemble from a
single trajectory were a long time in coming, the statistical mechanics of ensembles is an excellent fit with
Monte Carlo techniques.  Let us illustrate this idea for the example of a quartic oscillator, with the
Hamiltonian ${\cal H}_4 = (q^4/4) + (p^2/2)$.

{\bf Figure 1} shows two versions of quartic-oscillator Monte Carlo ensembles. In the first $(q,p)$ pairs come
from the random number generator {\tt rund(intx,inty)}, with energies up to $(q^4/4) + (p^2/2) = 5$, selected
from within the randomly accessed rectangle
$$
[ \ -\sqrt[4]{20} < q <+\sqrt[4]{20} \ ; \ -\sqrt{10}<p<+\sqrt{10} \ ] \ .
$$
Enlargement shows several line segments in the microcanonical distribution at top left, indicating correlation,
greatly reduced by discarding every third random number. The canonical distribution at the right, with the same
number of accepted points (10,000) was generated by accepting random choices in a larger rectangle, $-10<(q,p)<+10$ 
with probability $e^{-(q^4/4+p^2/2)}$ by accepting $(q,p)$ whenever ${\cal R}<e^{-(q^4/4+p^2/2)}$.  This example, like the
harmonic moments simulation, is interesting in that both of them point out shortcomings of {\tt rund(intx,inty)}.
Though the correlations are too small to see here at the scale of the microcanonical case, they are quite obvious
in the canonical ensemble sample at the upper right.  For most purposes this same {\tt rund(intx,inty)} generator
is perfectly adequate.

\begin{figure}
\includegraphics[width=2.5in,angle=-90.]{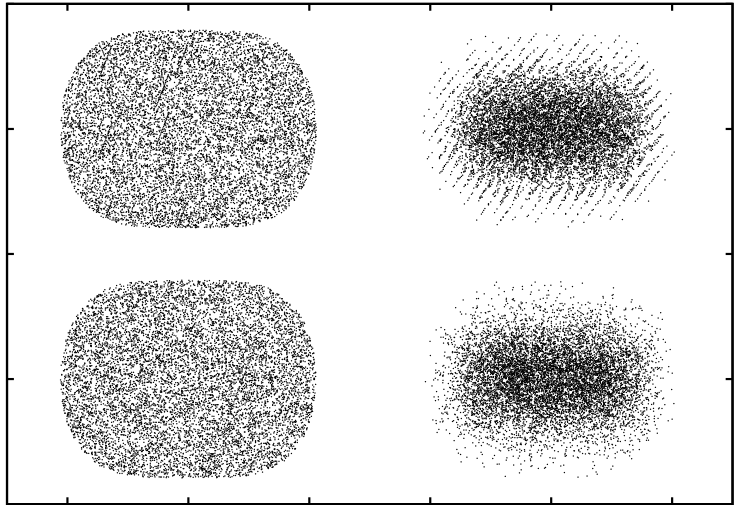}
\caption{
The random number generator {\tt rund} used to generate the ersatz microcanonical (at the left) and canonical
(at the right) distributions shows serial correlation in both cases. The flaws can be seen easily in an enlarged
view of this Figure.  Discarding every third of the random numbers improves the situation, as can be seen in the
canonical results to the right.  Each plot contains 10,000 points.
}
\end{figure}

\newpage

\section{Molecular Dynamics from the 1950s}

Gibbs' statistical mechanics is equilibrium-based.  Its simplest ``microcanonical ensemble'' formulation
describes many similar microscopic $\{ \ q,p \ \}$ copies ( the ``ensemble'' ) of a fixed mass of fluid
confined to a periodic volume $V$ with a fixed energy $E$. In the various copies the pressure and temperature
fluctuate, with their instantaneous values defined through the virial theorem and the ideal-gas
thermometer\cite{b13,b21}.  Gibbs' formalism describes the ``state'' of a manybody material as an average over
the ensemble of all applicable microstates. The usual optimistic alternative to the generating of such a
many-copy ensemble is to follow the ``molecular dynamics'' of a single ``typical'' specimen
system.  The 1950s began with Fermi, Pasta, and Ulam pursuing this idea at the Los Alamos Laboratory\cite{b18}.

Their chosen system was an anharmonic chain started with  a low-frequency longitudinal sinewave displacement.
Their motivating desire was to see how long it took for this ``typical'' anharmonic system to forget its
atypical initial condition, and to produce Gibbs' microcanonical equilibrium average properties.  They were
quite surprised to discover that anharmonicity was not enough to promote equilibrium. Their choice of problem
was therefore a good one. It has led to thousands of follow-on studies in the following seventy years.  Their
computational approach to the dynamics was likewise good.  Let us apply it to our harmonic-oscillator
example.

Fermi, Pasta, and Ulam used the second-order ``Leapfrog Algorithm'' to predict the next step in time from the two
preceding ones :
$$
\{ \ q(t\pm dt) \equiv q(t) \pm v(t \pm \textstyle{\frac{1}{2}}dt)dt \ ; \ 
v(t \pm \textstyle{\frac{1}{2}}dt) \equiv v(t \mp \textstyle{\frac{1}{2}}dt) \pm a(t)dt \ \} \longrightarrow
$$
$$
q(t+dt) = 2q(t) - q(t-dt) + (dt^2)a(t) \ [ \ {\rm Leapfrog} \ {\rm Algorithm} \ ] \ .
$$
The coordinate, velocity, and acceleration are respectively $(q,v,a)$.  Though the velocity appears in the
underlying ``leapfrog derivation'' the coordinates can be calculated as centered second differences without
any need to calculate or store the half-step velocities.

For the harmonic oscillator the acceleration is $-q(t)$ and the analytic solution of the finite-difference
algorithm is periodic in time, but with a slightly higher oscillation frequency deviating quadratically from the
exact oscillator trajectory, $q = \cos(t)$:
$$
q(t) = \cos(\omega t) \ ; \ \omega dt = \cos^{-1}[1 - (dt^2/2)] \
[ \ {\rm Leapfrog} \ {\rm Algorithm} \ {\rm Solution} \ ] \ .
$$
By a simple ``rescaling of the time'', a concept to which we return in Section V, this approximate algorithm can
be made exact for the oscillator. Fermi, Pasta, and Ulam used the leapfrog algorithm to study the dynamical
properties of anharmonic chains. They were mightily surprised that the chains showed no simple approach to equilibrium.

Soon after, Berni Alder and Tom Wainwright, helped by Mary Ann Mansigh (now Karlsen)\cite{b22}, at the Livermore
Laboratory in California, began to study the ``event-driven'' molecular dynamics of hard disks and spheres\cite{b2}. They
computed the times to each pair of particles' next collision accurately. The resulting geometry, coupled with
conservation of momentum and energy, gives the post-collision velocities of the two colliding particles. The
simulation then continues to the next collision.

Unlike Fermi's nonlinear chains, hard-particle systems soon came to thermal equilibrium, nicely reproducing the
Maxwell-Boltzmann velocity distribution $f(v) \propto e^{-mv^2/2kT}$. The most surprising finding of the Livermore
work was that (two-dimensional) hard disks underwent a fluid-solid phase transformation at a density near three-fourths
of the closest-packed ``triangular-lattice'' structure. The details of the transition have been progressively refined, as
recently as 2013, and are now quite well known\cite{b23}. Paradoxically the transition in three dimensions, with hard
spheres freezing at two-thirds the close-packed density, had already been characterized fairly well, by both molecular
dynamics and Monte Carlo, in the 1950s\cite{b19,b20}.  The three-dimensional transition is both broader and sharper than
is its two-dimensional little brother.

\newpage
 
\section{Isokinetic Nonequilibrium Dynamics from the 1970s}

\begin{figure}
\includegraphics[width=3in,angle=-90.]{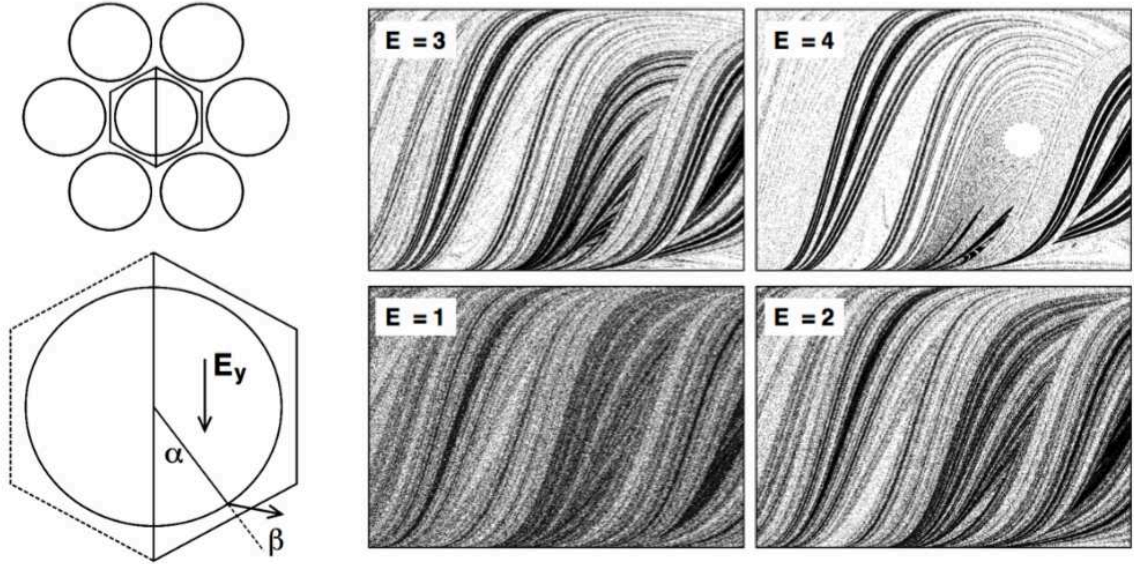}
\caption{
To the left we show the definition of $(\alpha,\beta)$ which define the location and exit velocity of each
Galton-Board collision.  To the right we see the fractal attractor distributions that result with gravitational
field strengths of $E = 1$, $2$, $3$, and $4$. This work is fully described in Reference 24.
}
\end{figure}

Molecular dynamics, when equipped with boundary conditions allowing heat transfer, can describe {\it dissipative}
stationary states requiring work for their maintenance and discharging an equivalent time-averaged amount of heat to their
environment.  Nonequilibrium molecular dynamics simulations with boundary velocities and temperatures were first
performed in the 1970s,\cite{b13,b21} by controlling particle velocities with ``thermostat forces''.  Implementing this
idea required a definition of ``temperature'', for which the ideal-gas definition, $mkT \equiv \langle  \ p^2 \ \rangle$
was readily adopted.  Gibbs' statistical mechanics had backed this definition with the observation that at equilibrium, so
long as the potential energy $\Phi(q)$ was independent of the kinetic, $K = \sum mv^2/2 = NDkT/2$. Here $D$ is the
dimensionality and $ND$ is the number of Cartesian degrees of freedom in the sum.  Gibbs' statistical mechanics
shows that exactly the same velocity distribution applies to dense matter as to the ideal gas.  These same thermostat
forces can also be generalized to simulating the mix of energy states required for Gibbs' canonical weighting of energies,
$e^{-E/kT}$.

The Department of Applied Science of the University of California at Davis was founded in 1963 in response to Edward
Teller's wish for a wider disemination of the research opportunities at the Lawrence Radiation Laboratory. By 1970 Berni Alder had helped
Bill Hoover to a Professorship there. Bill soon found a willing student, Bill Ashurst, from the Sandia Laboratory across
the street [ East Avenue ], to work with him on nonequilibrium simulation techniques.  Ashurst's PhD project,
{\it Dense Fluid Shear Viscosity and Thermal Conductivity via Nonequilibrium Molecular Dynamics}, was carried out
between January 1972 and May 1974. Ashurst developed computational ``fluid walls'', chambers containing a fixed number
of particles, with reflecting boundaries. At the end of each Leapfrog timestep the first and second velocity moments
in the fluid walls were adjusted to conform to the specified mean velocity and temperature:
$\{ v \rightarrow v + \alpha + \beta v \}$.

Bill's demonstration problems included dozens of simulations of viscous flows for dense fluids confined between two
oppositely-moving fluid walls. Offsetting the work done and the heat transferred by the elastic wall collisions and the
continuous momentum flow between wall and system particles, Ashurst maintained  nonequilibrium steady states by
alternating leapfrog steps with fluid-wall velocity adjustments on the order of a percent, maintaining the boundary
velocities and temperatures.  In this two-step process work was performed, and heat extracted, in such a way as to
obey a time-averaged version of the Second Law. 

Perhaps the simplest example problem constrains a manybody system to a fixed kinetic temperature.  To do this, using
a frictional force, $-\zeta p$, the constraint of fixed kinetic energy has the form :
$$
\{ \ \dot p = F - \zeta p \ \} \ ; \ \sum_1^N p \cdot \dot p = 0 = \sum_1^N \, [ \ F  - \zeta p \ ] \cdot p
\rightarrow \zeta = \sum_1^N F\cdot p /\sum_1^N p \cdot p \ .
$$
The thermostat forces $\{ \ -\zeta p \ \}$ exactly offset the natural fluctuations in the kinetic energy, forcing
it to remain constant, providing isothermal ( and isokinetic ) fluid walls.

It is interesting to see that the friction coefficient $\zeta$, being proportional to the momenta, is time-reversible
so that a reversed trajectory, with all of the $\{ \ p \ \}$ together with the two wall values of  $\{ \ \zeta \ \}$
changing signs, satisfies exactly the same motion  equations as did the forward trajectory.  This time-symmetry is
{\it paradoxical} ( Loschmidt's Paradox ) because {\it any} {\it irreversible} process, viewed backward in time, makes no
sense.

The physical explanation of the paradox was clarified in 1987\cite{b24,b25}. Bill, Harald Posch, Brad Holian, and another
PhD student of Bill's, Bill Moran, discovered that time-reversible thermostat forces obeying the Second Law of
Thermodynamics provide a dynamics which collapses onto a fractal ( fractional dimensional ! ) attractor.  These
attractors are made up of a negligible fraction of all the Gibbs' states present at equilibrium, but are still Lyapunov
unstable, with tiny changes in the conditions leading to exponentially growing differences in the future.  The
reversed trajectory, likewise containing its own negligible portion of phase space ( the ``repellor'', with reversed
velocities ) is much less stable than the attractor. The reversed trajectory is invariably less stable than the
attractor and cannot be generated directly from the dynamical equations.  The only way to obtain the reversed trajectory
is previously to compute, store, and reverse a forward trajectory.

One of the 1987 Toy Models\cite{b24}, the ``Galton Board'', demonstrated this explanation of the Second Law in terms of
isokinetic dynamics. A particle falling through a regular lattice of fixed scatterers, in the presence of a
gravitational field, was constrained to move at constant speed by imposing an isokinetic constraint on its
motion.  See {\bf Figure 2}. The locations of successive collisions with scatterers could be described by two angles.  The angle
$\alpha$ gives the location of the collision relative to the scatterer and the angle $\beta$ describes the
outgoing velocity direction after that collision. For hard-disk scatterers the equilibrium distribution of the
angles is perfectly uniform in a simple rectangle, with $0<\alpha<\pi$ and $-1<\sin(\beta)<+1$. The {\bf Figure}
shows the definitions of the two angles to the left and the distributions of collisions that follow from four gravitational
field strengths.  The fractal attractor dimension occupied by the collisional states decreases with increasing field
strength. Careful investigation shows that the dimension is always fractional so that the fraction of states in
the steady state is negligible, just as the measures of a line in two-dimensional space or a surface in three
dimensions have zero measure.

Ashurst's {\it heat}-flow simulations were carried out by maintaining two fluid walls at different isokinetic
temperatures. The resulting heat conductivities agreed fairly well with experimental data for liquid argon and
were used in simulating stationary shockwaves.  At Los Alamos this was accomplished by shrinking the volume as
a function of time.  At Livermore steady input and output flows far from the wave's center were used to generate
stationary shockwaves\cite{b26}.

\newpage

\section{The Influence of Nos\'e's 1984 Time-Scaling Dynamics}

Realistic computer simulations of dense fluid flows, even shockwaves\cite{b26}, caused the continuing explosion of interest
and participation in molecular dynamics we enjoy today.  With the background of the isokinetic and isoenergetic
nonequilibrium simulations of the 1970s many gifted researchers turned their attention to improving the state of
the art. Among them Shuichi Nos\'e was particularly innovative.  By an imaginative extension of Hamiltonian mechanics
he invented a method for mixing energy states dynamically in such a way as to reproduce Gibbs' canonical ensemble.
This linking of computer simulations to well-established fundamental physics helped popularize simulations and led to
the recognition of Berni Alder's pioneering influence with his award of the National Medal of Science in 2009.
Berni's award was followed four years later with the Nobel Prizes in Chemistry for Martin Karplus, Michael Levitt, and Arieh Warshel.
They developed realistic biological simulations using thermostated molecular dynamics with judicious
quantum-mechanical additions.

The isothermal mechanics invented by Shuichi Nos\'e in 1983-1984\cite{b3,b4} was a catalyst for this development. His
seminal work was greatly extended in 1984-1996\cite{b5,b6,b7,b9}. We introduce and discuss it here from a pedagogical
point of view. Replicating Gibbs' canonical ensemble with a deterministic Hamiltonian dynamics was Nos\'e's goal.
Two problems needed to be solved to accomplish it : [ 1 ] the new mechanics needed to access the energy states given
by Gibbs' canonical probability density, $f(q,p) \propto e^{-E/kT}$; [ 2 ] the new mechanics' phase-space trajectory
needed to speed up at higher energies and slow down at lower ones in just such a way as to convert the constant density
microcanonical distribution to the exponential density canonical one.  Nos\'e adopted Hamiltonian mechanics as his
starting point. With the imaginative addition of ``time scaling'' he could vary the speed of phase-space travel
(and more fundamentally the {\it strain-rate} of the corresponding compressible flow) to replicate Gibbs' distribution.

Nos\'e's highly original approach involves first the scaling of all the Hamiltonian momenta by a multiplicative
time-scaling factor $(1/s)\equiv e^{+E/kT}e^{+\zeta^2\tau^2/2}$, where $\zeta = p_s$ is a friction coefficient as well as
the Hamiltonian momentum conjugate to $s$, and where $\tau$ governs the strength of the thermostating forces
$\{ -\zeta\tau^2 p\}$.  Next, and finally, Nos\'e's fundamental invention, the time-scaling factor $s$, must adjust the
frequency of appearance of phase-space states in direct proportion to $e^{-E/kT} \propto f(q,p)$, Gibbs' phase-space
probability density.

Shortly after meeting with Nos\'e to discuss his work Hoover showed that frictional forces, $\{ \ -\zeta p \ \}$, where
$\zeta$ is determined by time-reversible feedback, $\dot \zeta \propto \sum (p^2 - mkT)$, reproduce the results of Nos\'e's
isothermal mechanics directly from the phase-space continuity equation\cite{b5}.

At Philippe Choquard's Lausanne laboratory Hoover applied Nos\'e's ideas to numerical simulations of a thermostated
one-dimensional harmonic oscillator.  This numerical work showed that the Nos\'e oscillator was far from ergodic. Instead, it
generated a remarkable variety of toroidal solutions as well as a relatively-small chaotic sea\cite{b6}.  The union of all
these separate solutions was the three-dimensional Gaussian :
$$
f(q,p,\zeta) = e^{-{\cal H}/kT}e^{-\zeta^2\tau^2/2} \equiv
e^{-[ \ q^2+p^2 \ ]/2}e^{-\zeta^2/2} \ [ \ {\rm Nos\acute{e}-Hoover} \ ] \ .
$$
A generation later Clint Sprott and the Hoovers showed that the oscillator model had toroidal orbits that formed
interlocking rings\cite{b27}.  More recently Lei Wang and Xiao-Song Yang found Nos\'e-Hoover oscillator trajectories in
the form of knots, very far from the simple ellipses of the isoenergetic model\cite{b28}. Here and in what follows, we
forgo those fascinating topological surprises, mercilessly simplifying Nos\'e's approach and its many possible
generalizations, as formalized by Bauer, Bulgac, and Kusnezov\cite{b14}.  Instead we choose to focus on the canonical
oscillator problem with linear friction and with all of the various parameters in the model, including $kT$, equal to
unity.

\subsection{Generalizations of Nos\'e-Hoover Mechanics, 1990-1992}

The oscillator-based discovery that Nos\'e's mechanics wasn't necessarily ergodic opened up a new research area which is
still quite active---finding motion equations which generate the entire canonical phase space distribution regardless of
initial conditions. The most useful work along those lines, with many worked-out example problems, was pioneered by Bauer,
Bulgac, and Kusnezov in two long and comprehensive readable papers in the Annals of Physics\cite{b14}. Their work generalized
Hoover's\cite{b13}, showing that several thermostat variables, called ``Demons'' in Reference 14, can be used simultaneously,
with three Demons enough to simulate one-particle  Brownian motion !  Generally they found that additional nonlinearity
enhances ergodicity. {\bf Figure 3} compares Poincar\'e sections at the plane $p = 0$ for three varieties of thermostated
oscillator including the far from ergodic Nos\'e-Hoover example :
$$
\{ \ \dot q = p \ ; \ \dot p = -q -\zeta p \ ; \ \dot \zeta = p^2 - 1 \ \} \rightarrow f =
e^{-q^2/2-p^2/2-\zeta^2/2} \ [ \ {\rm Nos\acute{e}-Hoover} \ ] \ ;
$$
$$
\{ \ \dot q = p \ ; \ \dot p = -q -\zeta^3p \ ; \ \dot \zeta = p^2 - 1 \ \} \rightarrow f =
e^{-q^2/2-p^2/2-\zeta^4/4}  \ [ \ {\rm Cubic} \ \zeta \ ] \ ;  
$$               
$$
\{ \ \dot q = p \ ; \ \dot p = -q -\zeta p^3 \ ; \ \dot \zeta = p^4 - 3p^2 \ \}  \rightarrow f  =
e^{-q^2/2-p^2/2-\zeta^2/2}  \ [ \ {\rm Cubic} \ p \ ] \ .  
$$

Before long, in 1996, Hoover and Holian found\cite{b29} that the simplest combination of two moment-based
Demons was enough to render the one-dimensional harmonic oscillator ergodic :
$$
\{ \ \dot q = p \ ; \ \dot p = -q -\zeta p - \xi p^3 \ ; \ \dot \zeta = p^2 - 1 \ ; \ \dot \xi = p^4 - 3p^2 \ \} \
[ \ {\rm Hoover-Holian} \ {\rm Ergodic} \ ] \ . 
$$
This set of equations provides an entire four-dimensional Gaussian distribution for any initial condition\cite{b29} :
$$
f = e^{-q^2/2-p^2/2-\zeta^2/2-\xi^2/2} \ [ \ {\rm Hoover-Holian} \ ] \ .
$$

In all of these cases the stationary distribution follows from the phase-space continuity equation, which is a
generalization of the ideas used to derive Liouville's Theorem :
$$                       
(\partial f/\partial t) = 0 = -f[(\partial \dot q/\partial q)+(\partial \dot p/\partial p) +                                               
(\partial \dot \zeta/\partial \zeta)] - \dot q(\partial f/\partial q) - \dot p(\partial f/\partial p)
-\dot \zeta(\partial f/\partial \zeta) \ .             
$$

\begin{figure}
\includegraphics[width=2.in,angle=-90.]{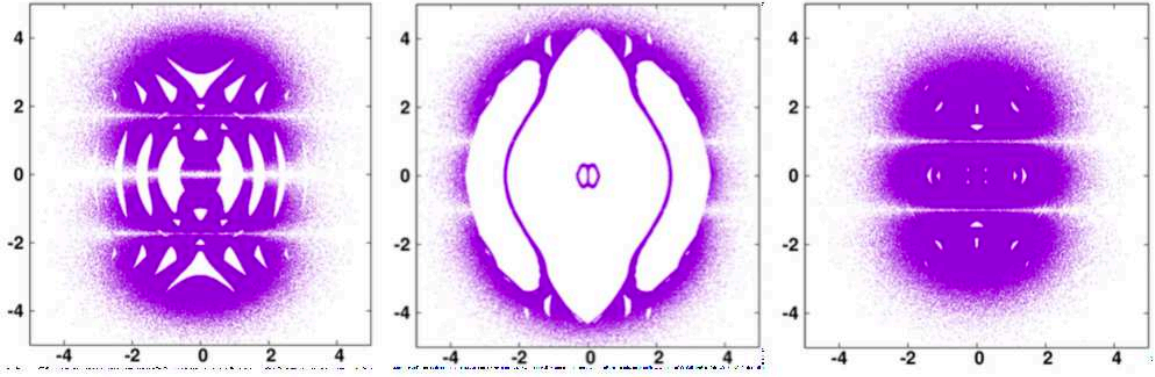}
\caption{
(q,p) cross sections for the Nos\'e-Hoover problem (center) with frictional force $-\zeta p$, as
well as two stiffer variations, $-p^3\zeta$ (left), and $-p\zeta^3$ (right) suggested by the work of
Hoover\cite{b13} and Bauer, Bulgac, and Kusnezov\cite{b14}.
}
\end{figure}

Although only the last of these approaches is ergodic recent developments have shown that a single thermostat
variable can provide ergodicity. A particularly interesting singular example was highlighted by Sprott\cite{b16}
as an extension of the prize-winning work of Tapias, Bravetti, and Sanders, who used a hyperbolic tangent function
of $\zeta$ to shift from heating (negative $\zeta$) to cooling (with positive $\zeta$)\cite{b15} :
$$
\{ \ \dot q = p \ ; \ \dot p = -q \mp \alpha p \ ; \ \dot \zeta = p^2 - 1 \ \} \
[ \ {\rm Sprott's} \ {\rm Signum} \ {\rm Thermostat} \ ] \ .
$$
Here the minus sign is used for positive $\zeta$ and the plus sign for negative $\zeta$. The momentum
$p$ varies continuously in time, but with an occasional discontinuity in its first derivative.  Sprott
observed ergodicity for this model provided that the parameter $\alpha$ is chosen at least equal to the
``Golden Ratio'', $1.618034 = \sqrt{(5/4)}+(1/2)$. The details of this work are not yet understood.
{\bf Figure 4} illustrates the uniform coverage of the Poincar\'e section $\zeta = 0$ obtained with Sprott's Thermostat.

\begin{figure}
\includegraphics[width=2.5in,angle=0.]{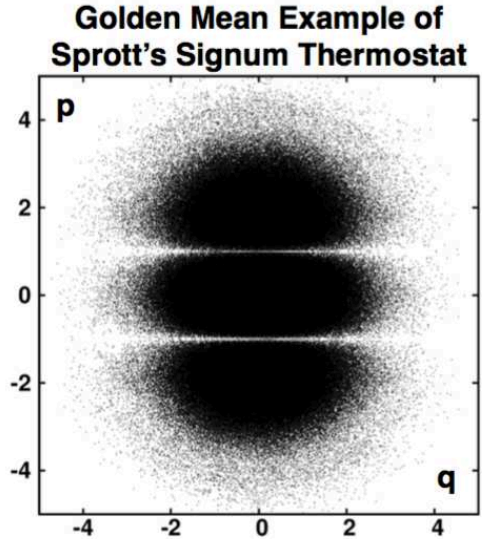}
\caption{
The cross-section $\zeta=0$ for the Signum Thermostat\cite{b16} with $\dot p = -q \mp 1.618034p$ for which the stationary
solution is $f(q,p) \propto e^{-[ \ q^2/2 + p^2/2 + 1.618304| \, \zeta \, | \ ]}$.  The ``nullclines'' at $p=\pm1$
show no penetration as the motion there is tangent to the Poincar\'e plane. For simplicity despite the
discontinuities in $\dot p$ we used 400,000,000 fourth-order Runge-Kutta timesteps with $dt = 0.0025$ to
approximate the distribution of $(q,p)$ points in the plane.
}
\end{figure}

\subsection{Dettmann's 1996 Contribution to an Understanding of Nos\'e's Approach}

Bill and Carl Dettmann discussed the difficulty of rationalizing Nos\'e's time-scaling step at a CECAM meeting
one July evening in Lyon\cite{b7,b9}.  By the next morning Dettmann had discovered that the simple step of multiplying Nos\'e's
Hamiltonian ${\cal H}_N$ by $s$, the mysterious time-scaling variable, provided a new Hamiltonian ${\cal H}_D$,
completely avoiding time scaling provided that this new Hamiltonian was chosen to have the value zero ! Dettmann's
Hamiltonian for the one-dimensional oscillator,
$$
{\cal H}_D \equiv s{\cal H}_N = s[ \ q^2 + (p/s)^2 + \zeta^2 + \ln(s^2) \ ]/2 \equiv 0 \ ,
$$
reproduces the Nos\'e-Hoover motion equations for the oscillator [ provided that the scaled momentum
$(p/s)$ is replaced by the symbol $p$ ].  As a fringe benefit, this step provides the identification of the
mysterious $s$ with the extended Gibbs' distribution $f(q,p,\zeta)$ ! :
$$
s \equiv f(q,p,\zeta) = e^{-[ \ q^2 + p^2 + \zeta^2  \ ]/2} \ [ \ {\rm Nos\acute{e}-Hoover = Dettmann} \ ] \ .
$$
Gibbs' three-dimensional Gaussian distribution can be converted into a probability density
for the time-scaling variable $s$ as follows:
$$
s^2 = e^{-(q^2 + (p/s)^2 + \zeta^2)} \equiv e^{-r^2} \rightarrow sds = -re^{-r^2}dr \rightarrow
(dr/ds) = -(1/rs) \rightarrow
$$
$$
prob(s) = (2\pi)^{-3/2}4\pi r^2e^{-r^2/2}(dr/ds) = \sqrt{(2/\pi)\ln(1/s^2)} \ [ \ {\rm Ergodic} \ ].
$$
To confirm this analysis we choose one million values of $r^2$ with the relative probability of
$r^2e^{1-r^2}$ and bin their logarithms in {\bf Figure 5}.

\begin{figure}
\includegraphics[width=2.5in,angle=+90.]{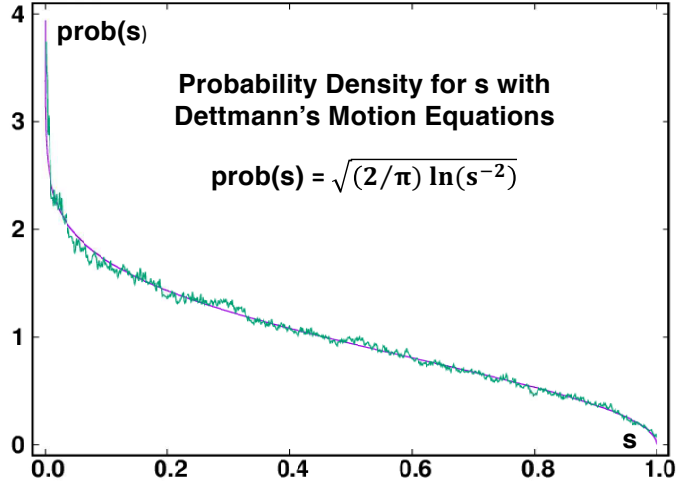}
\caption{
Probability density $prob(s)$ for $s=\sqrt{e^{-r^2}}$ where the radii $\{ \ r \ \}$ are selected from a
three-dimensional Gaussian distribution. One million data were sorted into one thousand bins and are
here compared with the analytic distribution derived in the text.
}
\end{figure}

In considering these extensions of Nos\'e's 1984 work we need to analyze a relatively simple model in order
to understand the relatively complex relationship between the speed of phase-space travel and probability
density.  The harmonic oscillator is too simple for this as the phase-space speed in entirely uniform. The
continuity equation for $f(q,p,\zeta)$ shows directly that rather than speed it is strain rate, $(\dot
\otimes/\otimes) \equiv (-\dot f/f)$ that is crucial to the equivalence between time scaling and probability
density. To clarify this point we consider the Quartic oscillator, with ${\cal H}_Q = (q^4/4)+(p^2/2)$.

\subsection{The Phase-Space Strain Rate of the Quartic Oscillator}

The equilibrium one-dimensional quartic oscillator, $\{ \ \dot q = p \ ; \ \dot p = -q^3 \ \}$, obeys Liouville's
Theorem,
$$
(\dot f/f) = -(\dot \otimes/\otimes) = - (\partial \dot q/\partial q) - (\partial \dot p/\partial p)
= -0 -0 = 0 \ .
$$
One might expect then that all the accessible oscillator states are equally likely, traversed at equal
speeds.  But they are not. The speed, with initial conditions $(q,p) = (0,1)$ varies between 1 and $2^{3/4}
= 1.6818$.  The oscillator conserves its energy so that its trajectory is just a one-dimensional line in
$(q,p)$ space, with a varying speed.  To the left in {\bf Figure 6} we see the oscillator $(q,p)$
trajectory and the time-dependence of the speed in phase space, $\sqrt{p^2+q^6}$.  To the right we see the
strain rate of the one-dimensional trajectory with initial condition $(q,p) = (0,1)$.  This is calculated
two ways: [ 1 ] the strain-rate parallel to the trajectory,  $(r\cdot v)/(r \cdot r)$ ; [ 2 ] the largest local
Lyapunov exponent, calculated by considering the constraint required to maintain the length of an
infinitesimal vector $(\delta_q,\delta_p)$ tied to the $(q,p)$ trajectory :
$$
\{ \ \dot q = p \ ; \ \dot p = - q^3 \ \} \rightarrow \{ \ \dot \delta_q = \delta_p - \lambda \delta_q \ ; \
\dot \delta_p = - 3q^2\delta_q - \lambda \delta_p \ \} \longrightarrow
$$
$$
 \lambda = \delta_q\delta_p(1 - 3q^2) \ .
$$
Because the motion is regular and periodic there can be no exponential growth of small perturbations.  But
the local values of the Lyapunov exponent vary in the range $[ \ -1 {\rm \ to \ } +1 \ ]$.
\begin{figure}
\includegraphics[width=2in,angle=-90.]{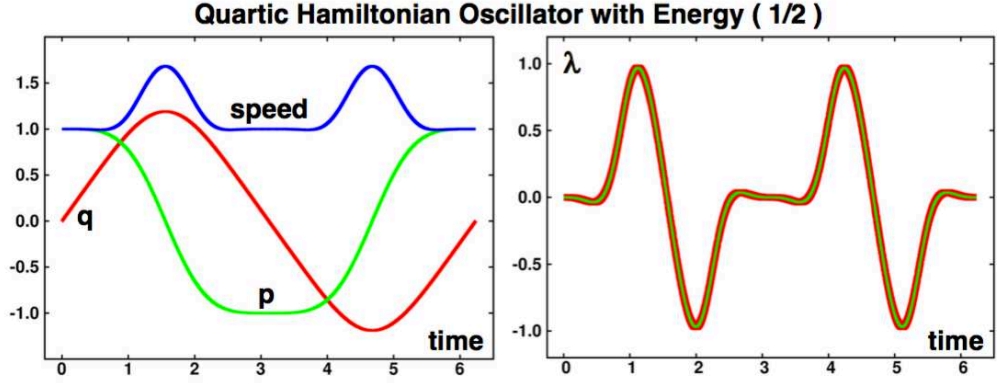}
\caption{
To the left we show the variation of $(q,p,\sqrt{(p^2 + q^6)})$ for one quartic oscillator period.  At the right
we show the variation of the one-dimensional strain rate along the $(q,p)$ quartic oscillator trajectory with
initial conditions $(q,p,\delta_q,\delta_p) = (0,1,1,0)$. The two methods mentioned in the text agree.
Note that the strain rate variation occurs twice during the oscillator period of approximately 6.236.
}
\end{figure}

This example shows forcefully that Liouville's Theorem is misleading when applied to a single-system
trajectory.  In the one-dimensional case, with only two phase-space directions, the longitudinal and
transverse strains exactly cancel, Liouville's Theorem. The transverse strain rate, eliminated by energy
conservation is equal to the second Lyapunov exponent.  The first, plotted to the right in {\bf Figure 6},
gives the local logarithmic rate of longitudinal expansion of an infinitesimal element of length as a function
of time. As the motion is periodic the mean value of both exponents is zero.  The analogous details for a
one-dimensional harmonic oscillator's elliptical phase-space orbit have been published in Reference 30.

\newpage

\section{An Apparent Nos\'e-Hoover-Dettmann Paradox}

In all there are three separate routes to exactly the same outcome, the Nos\'e-Hoover
oscillator motion equations and their stationary probability density :\\
$$
f(q,p,\zeta) = s(q,p,\zeta) = e^{-(q^2+p^2+\zeta^2)/2} \ :\\
$$ 
\noindent
[ 1 ] Nos\'e's Hamiltonian\cite{b3,b4}, followed by time-scaling ( multiplying all rates by $s$ ). \\
\noindent
[ 2 ] Hoover's continuity-equation derivation\cite{b5,b6} : $(-\dot f/f) = (\partial \dot q/\partial q) +
(\partial \dot p/\partial p) + (\partial \dot \zeta/\partial \zeta ) \ $. \\
\noindent
[ 3 ] Dettmann's Hamiltonian\cite{b7,b9}, set equal to zero and equivalent to setting Nos\'e's to zero. \\

Hoover's derivation has the simplest assumptions, depending only on the continuity of the variables $(q,p,\zeta,f)$,
separability, and linearity.  That is, assume that $f(q,p,\zeta)$ is separable, and of course positive, and that
$\zeta$ has the simplest possible (linear) effect on the trajectory. The consequence is a differential equation for
the functional dependence of the linear friction coefficient $\zeta$ on the oscillator variable $p$ :
$$
\{ \ \dot q = p \ ; \ \dot p = -q -\zeta p \ \} \ ; \ f = e^{-(q^2+p^2)/2}e^{g(\zeta)} \ ;
$$
$$
(\partial \ln f/\partial t) = 0 =  \dot qq + \dot pp - \dot \zeta (dg/d \zeta) - (\partial \dot p/\partial p)
= -\zeta p^2 - \dot \zeta (dg/d\zeta ) + \zeta \longrightarrow
$$
$$
\{ \ g = -(\zeta^2/2) \ ; \ \dot \zeta = p^2 - 1 \ \} \ .
$$
Alternatively, if the frictional force is cubic, $\dot p = -q - \zeta p^3$, we again find the Gaussian solution :
$$
0 = -\zeta p^4 - (p^4-3p^2)(-\zeta) + 3\zeta p^2\longleftrightarrow                                                                               
\{ \ g = -(\zeta^2/2) \ ; \ \dot \zeta = p^4 - 3p^2 \ \} \ .
$$
This alternative suggests that other odd powers of $p$ or other even integrable functions of $\zeta$ could be used
in its probability density, as is indeed the case\cite{b14,b15,b16}.

Let us next consider the difference between two descriptions of a thermostated oscillator---the one a
one-dimensional trajectory in a three or four-dimensional phase space; the other a three- or
four-dimensional flow of an ensemble of systems living in the same phase space.  We will focus on
the surprising qualitative differences among the three- and four-dimensional flows described by Nos\'e,
Dettmann, and Nos\'e-Hoover dynamics. All of them, even three-dimensional Nos\'e-Hoover, can be analyzed
in a four-dimensional $(q,p,s,\zeta)$ phase space, or in a three-dimensional subspace corresponding to
the single-trajectory restriction of constant energy. To promote the Nos\'e-Hoover flow to four dimensions
it is only necessary to define $\dot s \equiv s\zeta$. Liouville's Theorem then can apply to all three sets
of equations.  The Theorem establishes that the four-dimensional Hamiltonian probability density flows like
an incompressible fluid, with $\dot f \equiv 0$, just as in the familiar two-dimensional case :
$$
\{ \ \dot q = (\partial {\cal H}/\partial p) \ ; \ \dot p = -(\partial {\cal H}/\partial q) \ \} \longrightarrow
\dot f = (\partial f/\partial t) + \sum \dot q(\partial f/\partial q) + \dot p(\partial f/\partial p) \equiv 0 \ .
$$
The Nos\'e ($s^0$) and Dettmann ($s^1$) oscillator Hamiltonians differ by just a factor $s$ :
$$
{\cal H}_{N,D} = (s^{0,1}/2)[ \ q^2 + (p/s)^2 + \ln(s^2) + \zeta^2 \ ] \equiv 0 \ ;
 \ \zeta \equiv p_s \ .
$$
In both cases the resulting constant-energy dynamics develop in a three-dimensional constrained phase
space. For instance we can choose a space described by the coordinate $q$, scaled momentum $(p/s)$, and
friction coefficient $\zeta$.  With the energy fixed any one of the four variables $(q,p,s,\zeta)$ can
be determined from a convenient form of the constraint conditions :
$$
s = e^{ -(1/2)[ \ q^2+(p/s)^2+\zeta^2 \ ] } \ [ \ {\rm Dettmann} \ {\rm and} \ {\rm Nos\acute{e}} \ ] \ .
$$
It is convenient to specify $(q,p/s,\zeta)$ and then to select $s$ to satisfy the ${\cal H} \equiv 0$
constraints. A consequence of the Dettmann multiplier $s^1$ is the simple relationship linking solutions
of the Nos\'e and Dettmann Hamiltonians :
$$
(\dot q,\textstyle{\frac{d}{dt}}(p/s),\dot \zeta)_{Dettmann} \equiv
s(\dot q,\textstyle{\frac{d}{dt}}(p/s),\dot \zeta)_{Nos\acute{e}} \ .
$$
The Nos\'e and Dettmann trajectories are identical in shape but are traveled at different speeds.  Let us
illustrate the interesting differences among the three equivalent descriptions for the case of the simplest
periodic orbit. The initial conditions are (0, 0.46627, 0.30082, 0) so that initially the
scaled momentum is $(p/s) = 1.55$ and the Hamiltonian vanishes, with $s = e^{-1.55^2/2} = 0.30082$.

\subsection{An expanding model in four dimensions}

Nos\'e's Hamiltonian, ${\cal H}_N = (1/2)[ \ q^2 + (p/s)^2 + \ln(s^2) + \zeta^2 \ ]$, followed by
the time-scaling,
$(\dot q,\dot p,\dot s,\dot \zeta) \rightarrow (s\dot q, s\dot p ,s\dot s,s \dot \zeta)$,
 leads to four equations of motion in $(q,p,s,\zeta)$ space:
$$
\{ \ \dot q = (p/s) \ ; \ \dot p = -sq \ ; \ \dot s = s\zeta \ ; \
\dot \zeta = [ \ (p/s)^2 - 1 \ ] \ \} \ \rightarrow (\partial \dot s/\partial s) = +\zeta \
[ \ {\rm Dettmann} \ ].
$$
Exactly these same motion equations follow more simply from Dettmann's Hamiltonian, with no need of
time scaling. Because our initial condition has a higher ``temperature'', $(p/s)^2 = 2.4025$,
than the target of unity the short-time friction coefficient $\zeta$ becomes positive. This suggests,
from $\dot s = s\zeta$, that Nos\'e's (or Dettmann's ) oscillator's phase volume begins by {\it expanding}
rather than contracting. This expansion with a positive friction seems counter to Liouville's Theorem,
and suggests a paradox. {\bf Figure 7} shows the details of this four-dimensional problem. The time
scaling factor $s$ {\it is} precisely equal to Gibbs' canonical probability density. With the
short-time positive friction, $\zeta > 0$, the flow does contract rather than expand, despite the $\dot s$
equation. Let us investigate this intriguing problem further.

\begin{figure}
\includegraphics[width=2.5in,angle=-90.]{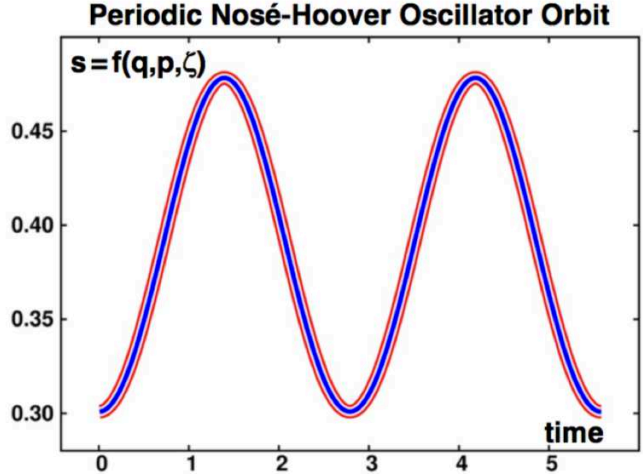}
\caption{
The time variation of three expressions for the probability density $f$ as measured once around a periodic
orbit generated with Dettmann's (or Nos\'e's, with time scaling) Hamiltonian in the four-dimensional
$(q,p,s,\zeta)$ phase space.  The initial conditions are (0, 0.46627, 0.30082, 0) so that initially the
scaled momentum is $(p/s) = 1.55$ and the Hamiltonian vanishes. The thickest line is Gibbs' canonical-ensemble
density chosen so that the initial value is $s= e^{-1.55^2/2}=e^{-[ \ q^2 + (p/s)^2 + \zeta^2 \ ]/2}$. The
medium white line overlaying the thicker red one shows the progress of the ``time-scaling factor'' $s(t)$.
The thinnest blue line is $s(0)e^{\int_0^t\zeta(t')dt'}$. The perfect agreement of the three demonstrates that
the phase-space density $f(q,p,\zeta)$ {\it can} be obtained by measuring the phase-space compression ( but not
the speed ) along the four-dimensional Hamiltonian trajectory with Dettmann's constraint, ${\cal H}_D \equiv 0$ .
But the early-time association of increasing phase volume, expected from
$(\partial \dot s/\partial s) = \zeta > 0$, is indeed paradoxical.
}
\end{figure}

\subsection{An incompressible model ?}

Dettmann's Hamiltonian, ${\cal H}_D = (s/2)[ \ q^2 + (p/s)^2 + \ln(s^2) + \zeta^2 \ ]$, with the constraint
${\cal H}_D\equiv 0$ imposed in the initial conditions, is not really {\it incompressible} :
$$
\{ \ \dot q = p/s \ ; \ \dot p = -sq \ ; \ \dot s = s\zeta \ ; \
\dot \zeta = -(1/2)[ \ q^2 - (p/s)^2 + \ln(s^2) + \zeta^2 \ ] - 1 \ \} \ \rightarrow
$$
$$
(\partial \dot s/\partial s) + (\partial \dot \zeta/\partial \zeta) = + \zeta - \zeta = 0 \
[ \ {\rm Incompressible?} \ ] \ .
$$

The flow equations certainly maintain a comoving {\it four}-dimensional hypervolume {\it unchanged in size}. This
is nothing more  than the usual application of Liouville's Theorem and is no surprise.  But taking the zero
energy constraint into account reduces the flow to three phase-space dimensions, just as in the Nos\'e-Hoover
picture.  Let us look at that picture next.  The quantitative details of the evolving phase probability are
shown in {\bf Figure 7}.

\subsection{A contracting model in three dimensions}

Here either Nos\'e-Hoover dynamics or a three-dimensional version of Dettmann's Hamiltonian,
including the constant-energy constraint, gives the same results. A time-reversible frictional
force, $-\zeta p$, provides a steady-state Gaussian phase-space distribution
$e^{-[ \ q^2+p^2+\zeta^2 \ ]/2}$.
In the two versions of dynamics the friction coefficient $\zeta$ is determined by the feedback
integral of temperature fluctuations around the target of unity :
$$
\{ \ \dot q = p \ ; \ \dot p = -q - \zeta p \ ; \ \dot \zeta = p^2 - 1 \ \} \ \longrightarrow
(\partial \dot p/\partial p) = -\zeta \ [ \ {\rm Nos\acute{e}-Hoover} \ ] \ .  $$

Dettmann's motion equations are identical to these if his scaled momentum $(p/s)$ is replaced by the
symbol $p$:
$$
\{ \ \dot q = (p/s) \ ; \ \dot p = -qs \ ; \ \dot s = s\zeta \ \} \ \stackrel{(p/s) \rightarrow p}{\longrightarrow} \ 
\{ \ \dot q =  p    \ ; \ \dot  p = -q - \zeta p \ ; \ \dot s = s \zeta \ \} \ .
$$
 Here, with the relatively ``hot'' initial condition, the three-dimensional phase-space
volume {\it shrinks} (correctly) initially due to contraction parallel to the momentum axis. So, for
the three phase-space descriptions of the same physical problem we have found expansion,
incompressibility, and compression, all for exactly the same phase-space states.  We put these three
examples forward from the standpoint of pedagogy, as a useful and memorable introduction to the
significance of Liouville's Theorem for isoenergetic flows. The constraint of constant energy can
lead to qualitative differences in the evolution of $f$ and $\otimes$.

\newpage

\section{Solving Differential Equations-Quantifying Ergodicity}

\begin{figure}
\includegraphics[width=2in,angle=-90.]{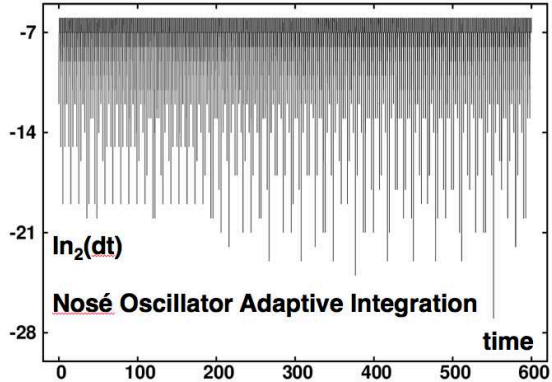}
\caption{
Variation of the timestep $dt$ required to bound the rms error, $\sqrt{dq^2+dp^2+ds^2+d\zeta^2}$
between $10^{-12} {\rm \ and \ } 10^{-10}$. With any error outside that range the timestep was
adjusted by a factor of two and the trial step was repeated. The initial conditions are taken
in the chaotic sea, $(q,p,s,\zeta) = ( 2.4,0,e^{-2.88},0 )$ and chosen so that the Nos\'e and
Dettmann Hamiltonians vanish.  The data shown correspond to about 250,000 successful timesteps.
}
\end{figure}

Mechanical simulations require solving differential equations and analyzing the results.
Solutions are necessarily numerical, almost always in the form of time series, and often
produced by packaged software.  Creating one's own software is both a pleasure and an
insurance policy, guarding against inflexible programming which is hard to understand
or improve.  Once the underlying model has been reduced to differential equations and once
these have been ``solved'', represented by a time series of salient variables ( coordinates,
momenta, energies, ... ), analysis takes over.  Once again it is simplest to maintain a
personal working library of transparent software for creating, displaying, and analyzing
data files.  In our own work there is a recurring need for the analysis of dynamical
instability, ``Lyapunov instability'', which causes small errors to grow, exponentially fast,
as $e^{\lambda(t)}$.  Let us consider numerical simulation work in more detail, beginning
with solving the equations and continuing with the analysis of the resulting data.

\subsection{Integration of Ordinary Differential Equations}
Although there is an extensive literature describing ``symplectic'' finite-difference schemes
for solving the molecular dynamics problems, much of it available on the arXiv, there is no real
need for these schemes in the research work we enjoy\cite{b17}.  In our experience fourth-order Runge-Kutta
integrators, where the programming is both simple and transparent, are best. Because  integration
errors vary as $dt^4$ over a fixed ( sufficiently short ) time interval, these can readily be
estimated by comparing the result of a single timestep $dt$, to the result of two timesteps of
half the length $(dt/2)$.

Let us define the integration ``error'' for a $dt$ step as the rms difference between the 
coarser $dt$ solution and the finer solution with two successive steps of $dt/2$. To illustrate
we consider Nos\'e's original Hamiltonian approach applied to the harmonic oscillator :
$$
\{ \ \dot q = p/s^2 \ ; \ \dot p = - q \ ; \ \dot s = \zeta \ ;
\ \dot \zeta = (p^2/s^3) - (1/s) \ \} \ [ \ {\rm Nos\acute{e}} \ ] \ .
$$
The rms error here is  $\sqrt{dq^2 + dp^2 + ds^2 + d\zeta ^2}$.

With double precision arithmetic it is convenient to choose $dt$ such that the error for the
oscillator lies between the values of $10^{-12}$ and $10^{-10}$. Whenever the error is too large,
greater than $10^{-10}$, we cut the timestep in half ; whenever the error is too small, less than
$10^{-12}$, we double $dt$.  Such an automated strategy is easily implemented and works quite well
with ``stiff'' differential equations like Nos\'e's or Sprott's Signum oscillator\cite{b17,b16}.

In {\bf Figure 8} we show the range of timesteps that results from these motion equations.
The integration error was constrained to lie between $10^{-10}$ and $10^{-12}$ for this demonstration.
One million successful steps were taken. The minimum step $2^{-28}$ lay below the average, $~2^{-9}$,
by about 19 powers of 2.  We show one quarter million steps in the figure.

\subsection{Achieving Ergodicity with the 0532 Model}

Gibbs' ensembles include {\it all} phase-space states consistent with the independent
thermodynamic variables, like energy, pressure, volume, and temperature. Particularly in
small systems with just a few phase-space dimensions dynamical ergodicity, as in the case
of the Signum thermostat of {\bf Figure 4}, is desirable. In 2015 it occurred to us that
``weak control'' could constitute a viable path to ergodicity.  This led us to the ``0532 Model'',
a smooth and ergodic representation of Gibbs' canonical distribution for the harmonic
oscillator\cite{b31}:
$$
\{ \ \dot q = p \ ; \ \dot p = -q -0.05\zeta p - 0.32\zeta (p^3/T) \ ; \
$$
$$
\dot \zeta = 0.05[ \ (p^2/T) - 1 \ ] + 0.32[ \ (p^4/T^2) - 3(p^2/T) \ ] \ \} \
[ \ {\rm 0532 \ Model} \ ] \ .
$$  
Notice that the model includes a linear combination of second-moment and fourth-moment controls
rather than one or the other of these possibilities.  There are many other combinations which
lead to ergodic dynamics.  For more details see Reference 31. The nullclines for the equilibrium
model, where $T=1$, are near $p = \pm 1.7$ where $\dot p$ vanishes.  Otherwise the $p(q)$ section
for the 0532 model looks much like Signum case of {\bf Figure 4}. Time and mirror symmetry for
the model imply fourfold symmetry in the Sections just as in the examples of {\bf Figures 3 and 4}.
Both of these symmetries disappear in the nonequilibrium case that the temperature becomes a
function of coordinate, the dynamics becomes dissipative, and the phase-space distribution becomes
fractal, all of which we illustrate next. The mirror symmetry is destroyed by the temperature
gradient while the time symmetry is destroyed by irreversibility, which allows for the dissipative
solutions that satisfy the Second Law of Thermodynamics, but steadfastly prevents their reversal.
See {\bf Figure 9} and note that both symmetries, $\pm q$ and $\pm p$ have disappeared.

\newpage

\section{Nonequilibrium Differential Equations and Maps}

\begin{figure}
\includegraphics[width=2in,angle=-90.]{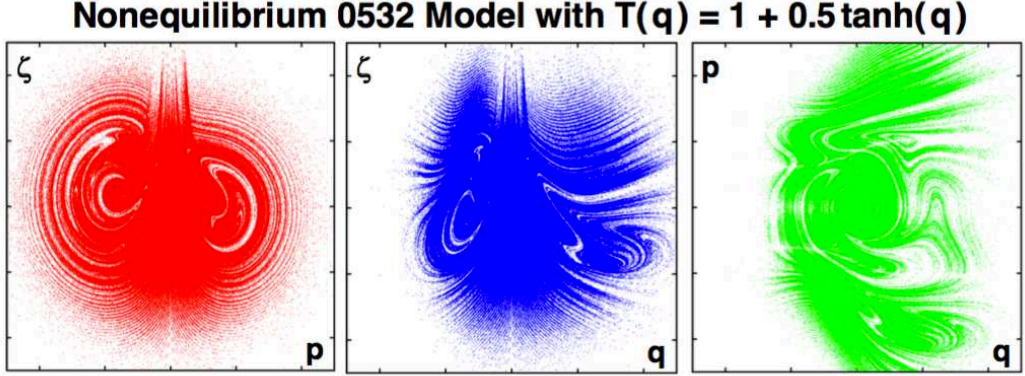}
\caption{
Three cross-sections of the 0532 model strange attractor with maximum temperature gradient of 0.5.
Each of the variables is shown in the range $-5 < \{ \ q,p,\zeta \ \} < + 5$ . The two nonzero
Lyapunov exponents for this system are +0.1135 and -0.1445, producing a ``strange attractor''.
The Kaplan-Yorke dimension of this fractal is 2 + (0.1135/0.1445) = 2.785, a zero-volume object in
the three-dimensional $(q,p,\zeta)$ space.
}
\end{figure}

Letting the temperature vary smoothly in the cold-to-hot range from 0.5 to 1.5 provides a simple
instructive three-dimensional nonequilibrium flow problem with a maximum temperature gradient of
0.5 :
$$
0.5 < T(q) \equiv 1 + 0.5\tanh(q) < 1.5 \rightarrow (dT/dq)_{q=0} = 0.5 \ .
$$
Although the {\it mass} current $\langle \ p \ \rangle$ necessarily vanishes, the {\it heat}
current  $Q = \langle \ p(p^2/2) \ \rangle$ does not.  Heat flows primarily from hot to cold and
is responsible for the dissipation which causes the collapse of phase volume ,
$$
\dot p = -0.05\zeta p - 0.32\zeta (p^3/T) \ [ \ {\rm 0532 \ Model} \ ] \longrightarrow
$$
$$
-(\dot S/k) = \langle \ (\dot \otimes/\otimes) = (\partial \dot p/\partial p) \ \rangle =
\langle \ -0.05\zeta - 0.96\zeta (p^2/T) = (Q/kT) \ \rangle = -0.0310 \ .
$$

This relatively simple example of a stationary nonequilibrium flow helps convey three 
lessons, treated in what follows : [ 1 ] Lyapunov instability, the exponential growth of
small perturbations. This is the main mechanism for the mixing of states; we will describe how to
characterize it. [ 2 ] The formation of strange attractors, with at least one positive Lyapunov
exponent but with a negative overall sum, is the typical situation away from equilibrium. The
irreversible attractors provide the microscopic analog of the macroscopic Second Law of
Thermodyamics. The simplest relevant example we know of is the time-reversible compressible Baker-Map
Model\cite{b32,b33,b34,b35}. [ 3 ] The fractal dimension of strange attractors can be related to
Lyapunov instability through the balance of chaotic growth with dissipative decay. We consider these
three lessons in turn, beginning with a look at Lyapunov's ideas from a bit over 100 years ago.

\subsection{Lyapunov Instability}

Alexander Lyapunov (1857-1918) characterized the (exponential) instability of differential
equations in terms of the growth and decay exponents describing the deformation of a phase
space hypersphere. An $N$-dimensional problem with $N$ ordinary differential equations is
characterized by $N$ exponents. Their sum gives the change of phase volume, $\sum \lambda =
(\dot \otimes/\otimes)$, zero at equilibrium and negative for nonequilibrium steady states,
corresponding to the formation of a strange attractor.

The simplest nonequilibrium flow problems are three-dimensional, the minimum for chaos.  They
can be described by three exponents, $\{ \ \lambda \ \}$ with the first and largest easy to
calculate, the second equal to zero, and the third negative, large enough to provide the
negative sum, $\lambda_1 + \lambda_3 < 0$ consistent with the Second Law of Thermodynamics.
The first and largest exponent, $\lambda_1 = \langle \, \lambda_1(t)\, \rangle$, can be
determined by measuring the growth rate of small perturbations.  In practice this is done by
following two neighboring solutions (a ``reference'' and its ``satellite'') and evaluating their
short-term tendency to separate. At the end of each timestep the separation $\delta$
is compared to the target value $\delta_0$, (typically $10^{-5} {\rm \ or \ } 10^{-6}$).  The
rescaling operation necessary to return the separation to the target value defines the local
exponent $\lambda_1(t)$ :
$$
(q,p,\zeta)_s = (q,p,\zeta)_r + (\delta_0/|\,\delta\,|)\,\delta \ ;
$$
$$
\delta \equiv[ \ (q,p,\zeta)_s - (q,p,\zeta)_r \ ] \ ; \
\lambda_1(t) \equiv -\ln(|\, \delta \, |/\delta_0)/dt \ .
$$
The third (negative) exponent gives the overall negative sum, $\lambda_1 + 0 + \lambda_3 < 0$,
required for convergence of the phase-space distribution. A straightforward method for finding
$\lambda_3$ when the equations are time-reversible, is to store and reverse a forward
trajectory\cite{b36}.  When analyzed backward (again keeping a satellite trajectory close to the
reversed reference) the largest Lyapunov exponent is $-\lambda_3$ and the negative exponent is
$-\lambda_1$.  The reversed trajectory, the ``repellor'', acts as a source for the phase-space
flow, from the repellor to the attractor.

In order better to understand the strange attractors we describe and illustrate a two-dimensional
map. This map conveys the same lessons as a three-dimensional flow, or a many-body dissipative
simulation, but in the simplest setting possible.  The two-dimensional map can be pictured as
relating two successive cross-sections of a three-dimensional flow.

\subsection{The Nonequilibrium Time-Reversible Compressible Baker Map}

The ``Baker Map'' name recalls the physical mixing implemented by kneading dough.  This pedagogical
nonequilibrium map\cite{b32,b33,b34,b35} ``N'' allows for the variable compression of the dough,
leading to a ``fractal'' (fractional-dimensional) loaf and to irreversible dissipation despite the
perfect time-reversibility of the underlying linear  equations.  The N mapping\cite{b34,b35} at the
left of {\bf Figure 10} is this :\\
For twofold expansion ( of the black region ), $q < p - \sqrt{2/9}$ :
$$
q^{\prime} = (11q/6)-(7p/6) + \sqrt{49/18} \ ; \ p^{\prime} = (11p/6) - (7q/6) - \sqrt{25/18} \ .
$$
For twofold contraction ( of the white region ),  $q > p - \sqrt{2/9}$ :
$$
q^{\prime} = (11q/12)-(7p/12) - \sqrt{49/72} \ ; \ p^{\prime} = (11p/12) - (7q/12) - \sqrt{1/72} \ .
$$
The mapping, $(q,p)\rightarrow(q^{\prime},p^{\prime})$, applies within a rotated $2\times2$ square
with extreme values of $q$ and $p$ of $\pm \sqrt{2}$. The ``T'' time-reversal mapping shown in
{\bf Figure 10} changes the sign of the ``momentum'' $p$, leaving the ``coordinate'' $q$ unchanged.
This diamond-shaped version of the map has the twin advantages of [ 1 ] time reversibility and
[ 2 ] square roots.  These roots circumvent the very short limit cycles which occur within the
simpler-looking but less useful ``square'' version of the same map, with $0<x,y<+1$ :
$$
2/3<x<1 \longrightarrow x^{\prime} = 3x - 2 \ ; \ y^{\prime} = (1+2y)/3 \ ;
$$
$$
0<x<2/3 \longrightarrow  x^{\prime} = 3x/2 \ ; \ y^{\prime} = y/3 \ .
$$
Single- and double-precision iterations of the $(x,y)$ map, starting at $(0.5,0.5)$ produce periodic
orbits of lengths, 1571 and 146,321,810, too short for statistical analyses.
A single-precision FORTRAN program of the diamond-shaped ``N'' mapping using the gnu compiler produced
a periodic orbit of 1,124,069 discrete $(q,p)$ points.  A double-precision iteration of this problem,
starting with $(q,p)=(0,0)$, showed no periodicity during $10^{12}$ iterations.  See pages 16-23 of
Lecture 9 and Section 3 from Lecture 10 of our Kharagpur Lectures vugraphs for more details. All eleven Lectures can be found at williamhoover.info on the web.

\subsection{Lyapunov Exponents for the Nonequilibrium Baker Map}

At the top left of {\bf Figure 10} a small element of white area expands by (3/2) and contracts by (1/3)
while a black element expands by 3 and contracts by (2/3). The inexorable resulting stretching in the
northwest-southeast direction leads to (2/3) of the measure white and (1/3) black.  These considerations
give for the longtime-averaged expansions and contractions :
$$
\lambda_1 = (2/3)\ln(3/2) + (1/3)\ln(3) = (1/3)\ln(27/4) = +0.63651 \ ;
$$
$$
\lambda_2 = (2/3)\ln(1/3) + (1/3)\ln(2/3) = (1/3)\ln(2/27) = -0.86756 \ .
$$
Thus a small one-dimensional line exposed to the mapping grows as $e^{0.63651t}$ with $t$ iterations
of the map while a tiny two-dimensional area shrinks as $e^{-0.23105t}$. Kaplan and Yorke\cite{b32}
provided a simple approximate estimate, thought to be exact in this case, relating fractal structure to Lyapunov
instability, the third nonequilibrium lesson we'll relate to the Baker Map and conducting oscillator
models.

\begin{figure}
\includegraphics[width=2in,angle=-90.]{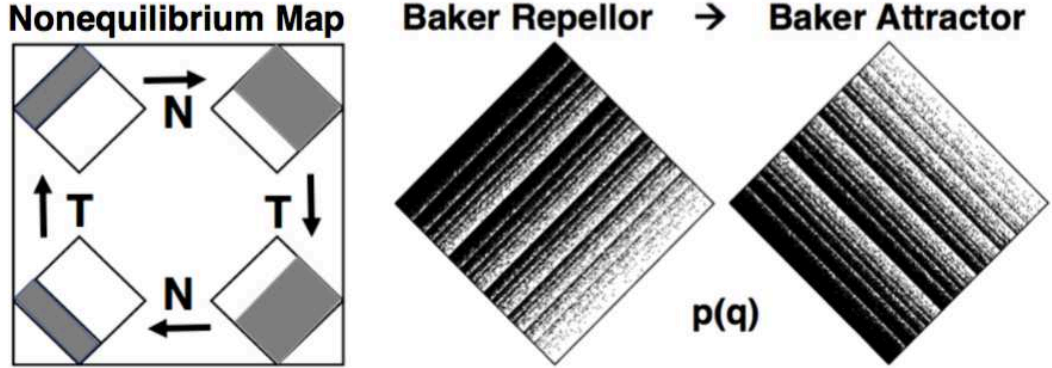}
\caption{
The nonequilibrium Baker Map ``N'' carries the southeast two-thirds at the left to the southwest
one-third at the right of the figure.  The flow from the repellor, at the center of the figure, to
the attractor at the right can only be reversed by storing and reversing (``T'') a forward trajectory.
The repellor has zero probability, with two-thirds of that in the northwest third at the left.  The
attractor has unit probability, with two-thirds of that in the southwest third at the right.
}
\end{figure}

\subsection{Fractal Dimensionality}
Dimensionality is simply related to scaling relationships.  A three-dimensional cube doubled
in size contains eight times the mass.  A square four times and a line two times.  The same
idea can be applied to fractional noninteger ideas about dimensionality.  If a steady-state
structure in phase space has growth and decay rates of +0.1135 and -0.1445 respectively the
area covered varies as $e^{-0.0310t}$ while the length varies as $e^{0.1135t}$.
Characterizing ergodicity in a three-dimensional problem can be addressed by sectioning the
phase space in a search for (nonergodic) ``holes''.  See again the $(q,p)$ sections of
{\bf Figure 3}.

Despite the additional complexity of the rotated coordinate system the $(q,p)$ version has
the physical advantage of time reversibility and the computational advantage of irrational
numbers in the mapping, which substantially delay the formation of periodic orbits.

The information dimension $D_I$ is the limiting small-mesh ratio  $\langle \, \ln(p) \,
\rangle /\ln(\delta)$ where $p$ is the probability associated with an element of the mesh
and $\delta$ is the mesh size.  Evidently a $D$-dimensional object of unit volume would have
mesh element probabilities of $\delta^D$ and the resulting average would agree with the
ordinary notion of (integer) dimensionality.  To analyze the Baker Map the simplest approach
is to store a reasonable number of $(x,y)$ points, $2\times 10^{11}$ rotated $(q,p)$ points,
reduced to lie within the unit square. For illustrative purposes we use two hundred billion
successive $(x,y)$ points. This takes a few hours' effort on a typical laptop computer.  The
information dimension was conjectured by Kaplan and Yorke to agree with a linear interpolation
to zero strain rate between the last positive Lyapunov sum (starting with the largest value)
and the first negative sum.  For the Baker Map in two dimensions this gives the estimate:
$$
D_I \stackrel{?}{=} D_{KY} \equiv 1-(\lambda_1/\lambda_2) =
1 + \ln(27/4)/\ln(27/2) = 1.733680 \ .
$$

Although the Kaplan-Yorke conjecture is plausible ( estimating the blend of expansion and
contraction which gives a vanishing strain rate ) and has been proved true\cite{b32,b33} for
a wide variety of maps, there are examples in which it definitely appears to fail\cite{b37}.
The pedagogical simplicity of the Map suggests it as a canonical analog of nonequilibrium
simulations, fit for numerical and theoretical exploration.  A plausible statistical model
follows from the observation ( easily verified numerically ) that two-thirds of the Map
iterations give compression in the $y$ direction and one-third give expansion. This Map is
therefore equivalent to a one-dimensional random walk with a variable step length. Choosing
a random number ${\cal R}$ for each iteration the stochastic model we use ( for $0 < y < 1$ )
is :
$$
{\cal R} < 2/3 \rightarrow y = y/3 \ ; \
{\cal R} > 2/3 \rightarrow y = (1+2y)/3 \ . 
$$
The information dimension for as many as a trillion iterations of the FORTRAN {\tt Random\_Number}
routine can then be analyzed using a mesh length $(1/3)^n$ for $n$ as large as 19. A plot
of some easily accessible results, $D_I$ as a function of $-1/\ln(\delta)$, provides a nice straight
line, as shown in {\bf Figure 11}.  An apparent fly in the Kaplan-Yorke ointment can be seen by
looking at the Kaplan-Yorke estimate as $\delta$ approaches zero.  $D_{KY} \simeq 0.7337$ while our
numerical estimate from data is $0.74_{15} \pm 0.001$.  Thomas Gilbert pointed out to us that the
convergence of the information-dimension calculation can be nonuniform. In the fine-mesh limit
{\it both} the number of iterations and the number of bins must be large.  He suggested that a
different mesh, $\delta = 2^{-n}$ rather than $\delta = 3^{-n}$ might give quite different results.
We found this to be true, making computational determinations of the information dimension somewhat
problematic for nonlinear problems.

Our numerical random-walk results for $D_I$ conform to theory, agreeing with the $(q,p)$ mapping
``information dimensions'' for feasible
values of $\delta$.  Despite this agreement it is true that the nonuniform convergence of the
limiting process means that the extrapolation of {\bf Figure 11} is incorrect!
$$
D_I \equiv 1 + \sum_i p_i\ln(p_i)/\ln(1/\delta) \ [ \ {\rm one \ dimension} \ ] \ . 
$$
Such determinations are much more economical than their two-dimensional twins :
$$
D_I \equiv \sum_i\sum_j p_{i,j}\ln(p_{i,j})/\ln(1/\delta) \ 
[ \ {\rm two \ dimensions} \ ] \ .
$$ 
Our detailed investigation of this Baker Map dimensionality took us a few days and is still
under investigation. Evidently this project was well worthwhile.  The results so far suggest
compressible Baker Map estimates from the statistical model agree with those using 
two-dimensional meshes.  It was a surprise to find that the limiting $\delta \rightarrow 0$
$D_I$ given correctly by Kaplan-Yorke is prone to error when pursued by systematic extrapolation.
Further unpublished results with trillions of iterations and $n=19$ are fully consistent with the
figure.  We expect to report more details on this interesting feature of the compressible Baker
Map.

\begin{figure}
\includegraphics[width=4.0in,angle=-90.]{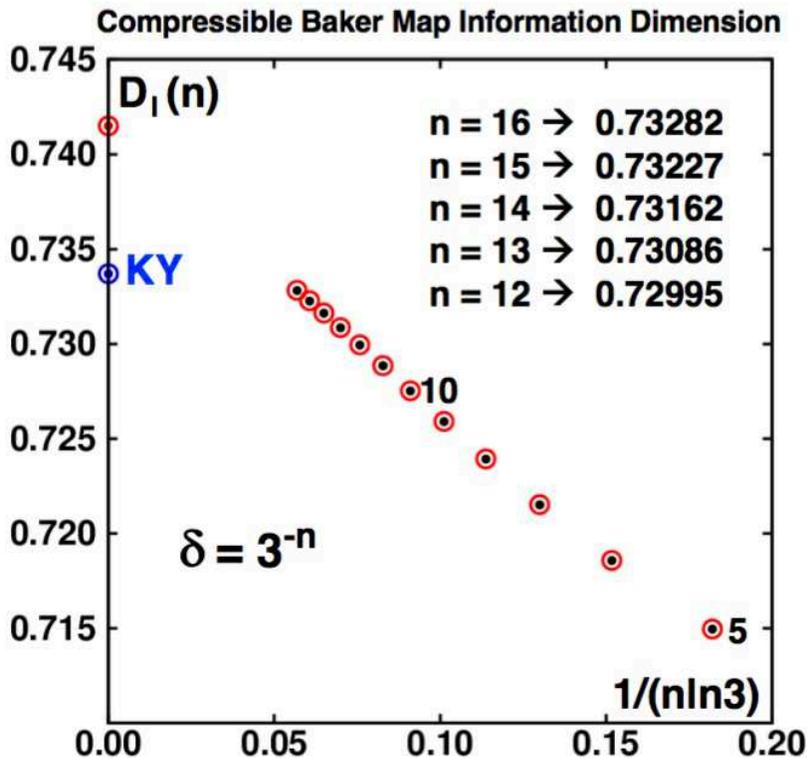}
\caption{
Variation of the apparent information dimension of the unit-square Baker Map with the mesh size
$\{ \, 3^{-n} \, \}$, with $4 < n < 17$ and $2\times 10^{11}$ iterations of the $(q,p)$
map analyzed in the unit $(x,y)$ square. The random-walk model, with compressive steps
for $0 < {\cal R} < 2/3$ and expanding steps for $2/3 < {\cal R} < 1$, and the complete
two-dimensional mapping are compared with the red and black points.  The two approaches are
consistent with each other to five figures and suggested incorrectly that the Kaplan-Yorke
information dimension ( blue ) was incorrect.  The dependence of the apparent value of $D_I$ on
the choice of mesh was a surprise and deserves more attention.
}
\end{figure}

\newpage

\section{Irreversibility and the Second Law of Thermodynamics}

The agreement of the linear interpolation and stochastic models to the deterministic 
time-reversible Baker Map provides an intuitive understanding of the Second Law of
Thermodynamics\cite{b25}. In the general case of a thermostated time-reversible
nonequilibrium steady state the longtime-averaged flow {\it from} a zero-probability
fractal repellor {\it to} its mirror-image zero-volume strange attractor is invariably
``dissipative''.  Microscopic dissipation is characterized by phase-volume shrinking,
$\otimes \rightarrow 0$  as the macroscopic dissipative heat is generated by the flow
and extracted by time-reversible ``thermostated'' motion equations.

It is because the repellor has a positive Lyapunov sum, corresponding to an (impossible)
exponential divergence of the phase volume, $\otimes \rightarrow \infty$, that repellor
states can only arise by storing and reversing a dissipative trajectory.  With digital
computers any stationary state simulation will eventually generate a periodic orbit\cite{b38}.
Using single-precision arithmetic the Baker Map generates a periodic orbit of reasonable length.
With double precision the length of the period is inaccessible to an informal investigation.

We believe that the most valuable result catalyzed by Nos\'e's exploration of computational
thermostats is the understanding of the inevitable statistical favoring of flows obeying the
Second Law (that nonequilibrium flows are dissipative). Likewise one can simply look and see
that nonequilibrium states are of vanishing probability relative to Gibbs' equilibrium states.
The fascinating fractal character of nonequilibrium states underlines the interest in the
topological study of phase-space structure.  One can imagine a continuous probability
distribution becoming fractal. This picture seems entirely unlike the one-dimensional 
trajectory pursued by an individual nonequilibrium system in its point-by-point exploration
of $6N$-dimensional floating-point phase space. Whether or not the distinction between [ 1 ]
continuous variables and [ 2 ] the digital ones we use in modeling them is significant could
use a transparent investigation from a kind-hearted mathematician, assuming his existence!

\newpage

\section{Future Characterizations of Dynamical Systems}
The oscillator, Galton Board, and Baker Map problems provide an excellent introduction to nonlinear
dynamics, chaos, and, by venturing into the many nonequilibrium applications of the thermostat idea,
fractal geometry. A particularly simple nonequilibrium model is the one-dimensional $\phi^4$ chain,
where each particle interacts with a harmonic nearest-neighbor force and is also tethered to its
lattice site by a quartic potential.  Adding thermostat forces to both ends of the chain results in
a conductive heat current from the ``hot'' end to the ``cold'' one and invariably provides a fractal
( fractional-dimensional ) phase-space attractor\cite{b39}.  See again {\bf Figure 9} for the
oscillator version of such a fractal.

The generic properties of the compressible Baker Map [ Lyapunov instability, steady-state
irreversible flow {\it from} a zero-volume ergodic fractal repellor {\it to} a mirror-image
strange attractor, quadratic dependence of the dissipation rate, $(\dot \otimes/\otimes)$,
on the deviation from equilibrium ] provide a fine illustration of the macroscopic
Second Law of Thermodynamics in terms of a microscopic time-reversible deterministic
thermomechanics. At the same time the fractal nature of the strange attractor-repellor pair
still contains mysteries appropriate to more computational research. Mathematics seems to be
of little help here. The very notion of an attractor in mathematics seems qualitatively
unlike our computational observations.

In mathematics an attractor is thought of as an ``infinite'' set of points, but with the concept
of infinity muddled by the undecidability of the continuum hypothesis. The concept of the
cardinal number $\aleph_0$ as the number of integers, or rationals, is not at all controversial.
That a ``continuum'' is different is obvious so that a count of points ``in'' the continuum
introduces a new infinity, sometimes called $c$. G\"odel is credited with showing that it can't
be shown whether or not $c$ and $\aleph_1 = 2^{\aleph_0}$ are one and the same. This standstill
has lasted nearly a century.  At the moment the validity of the continuum hypothesis looks
suspiciously like a ``meaningless question'', divorced from the reality of computation.

From the computational standpoint it appears that our floating-point numbers, all of them
rational, are certainly not a continuum.  But they represent it well.  Even quadruple-precision
arithmetic (closer to the continuum ?) is tedious in practice and typically teaches us nothing new.
From the computational standpoint the number of points in a two-dimensional array can be arbitrarily
larger than the number of those  in a one-dimensional array. Likewise for three dimensions relative
to two.  In mathematics whether the continuum is one-, two-, or three-dimensional the ``number of
points'' $c$ is all the same.  It is here that mathematics seems to deviate from useful to useless.

The computational analysis of fractals introduces a nonintegral dimension missing from
mathematics.  In the vicinity of a point within a fractal one can characterize the density of
nearby points with a power law, $\delta^D$. $D$ can be nonintegral--the existence of the power
law can vary wildly with direction and can be made more precise and detailed by increasing the
precision or decreasing the mesh size to the limit of one's budget. Example problems shedding
more light on the microstructure of nonequilibrium fractals remain a pressing need.

\section{Acknowledgment}

We specially thank Thomas Gilbert for pointing out that the apparent information dimension of
the data in {\bf Figure 11}, $D_I \rightarrow 0.741$, is incorrect and that a different choice of mesh,
$\delta = 2^{-n}$, gives yet another incorrect apparent dimension. For more details see our arXiv
report 1909.04526, ``Random Walk Equivalence to the Compressible Baker Map and the Kaplan-Yorke
Approximation to its Information Dimension''.  We specially appreciate Gilbert's
preparation of an unpublished Note detailing the statistical analysis of the compressible dissipative
Baker Maps.  Thomas, together with Carl Dettmann, Ed Ott, Harald Posch, and James Yorke, also pointed
us to the literature\cite{b32,b33} establishing that the Kaplan-Yorke conjecture has been proved true
for linear Baker Maps of the type illustrated in {\bf Figures 10 and 11}.

\end{document}